\documentclass[]{tPHL2e}

\newcommand \OMIT[1]{}
\newcommand \SAVE[1]{}
\newcommand \MEMO[1]{{\bf #1 }}

\newcommand{\Mg}{_{\rm Mg}}
\newcommand{\Zn}{_{\rm Zn}}
\newcommand{\la}{\langle}
\newcommand{\ra}{\rangle}
\newcommand{\house}{{\rm house}}
\newcommand{\MgMg}{_{\rm Mg-Mg}}
\newcommand{\MgZn}{_{\rm Mg-Zn}}
\newcommand{\ZnZn}{_{\rm Zn-Zn}}

\newcommand{\HHtile}{{\cal H}_{\rm tile}}
\newcommand\mgivznvii{Zn$_7$Mg$_4$}

\newcommand\FrH{F/H}

\newcommand\rt{T/R}

\begin{document}

\markboth{Decagonal Frank--Kasper Mg-Zn-RE}
{Mihalkovi\v{c}, Richmond-Decker, Henley, and Oxborrow}

\title{Ab-initio tiling and atomic structure for decagonal ZnMgDy quasicrystal}

\author{M. Mihalkovi\v{c}~\thanks{$^\dagger$
Corresponding author. Email: clh@ccmr.cornell.edu}$^\dagger$
Institute of Physics, Slovak Academy of Sciences, Dubravska cesta 9,
84511 Bratislava, Slovakia
\\\vspace{6pt} 
J. Richmond-Decker and C.~L.~Henley,
Dept. of Physics, Cornell University, Ithaca NY 14853-2501 USA
\\\vspace{6pt} 
M. Oxborrow, ...
National Physical Lab, ... UK
\received{-- --, 2011}}

\maketitle

\begin{abstract}
{
We discover the detailed atomic structure of $d$-MgZnY, 
a stable decagonal quasicrystal alloy of the layered Frank-Kasper
type, and related phases, using the ``tiling and decoration'' approach.
The atoms have invariable sites in the rectangle and triangle 
tiles of a 10-fold-symmetric planar tiling.
To discover the lowest-energy structures, we combine the methods of
density functional theory (DFT) total energy calculations,
empirical oscillating pair potentials (fitted to DFT),
fitting effective Hamiltonians for tilings,
and discovering  optimum tiling structures using a 
nonlocal tile-reshuffling algorithm.
We find a family of practically stable compounds with varying composition,
including the decagonal quasicrystal and the known Mg$_4$Zn$_7$ phase 
\SAVE{(which is the most stable);
these are more stable than competing icosahedral structures by a small margin.}
}
\OMIT{``We study the stability of all phases in the Mg--Zn system around the composition
Mg$_{40}$Zn$_{60}$, focusing on structures related to the decagonal quasicrystal.
Finally, our grand result is DFT--based zero--temperature energy diagram, 
spanning whole range of Mg--Zn compositions.''}
\bigskip
\bigskip
\end{abstract}


\section{Introduction}

\MEMO{CLH has downplayed all parts that emphasize the grand Mg--Zn $T=0$
phase diagram, and has parked that text in appendix \ref{sec:grand-Mg-Zn}.}

Quasicrystals are grouped into several major classes
according to their chemistry and local order, one of which
is the ``Frank-Kasper'' class. In such a structure~\cite{FK},
the atoms are packed such that neighbors always form tetrahedra,
\SAVE{(This follows from the reasonable assumption that
every interatomic bond is surrounded by either five or six atoms.}
and only four shells of neighbors are possible, having coordination
numbers CN=12, 14, 15, or 16.  
The Frank-Kasper structures in the decagonal phase are packings of 
atoms of two different sizes, such that ``small'' atoms 
have icosahedral (CN=12) coordinations, ``large'' atoms occupy 
the CN=15 or 16 sites, and either kind might occupy the CN=14 sites; 
the fractional content of $L$ atoms is typically 35--40\%. 
The same local order is found in well-known icosahedral Frank-Kasper 
quasicrystals, including stable $i$-AlMgZn \MEMO{CITE i-AlMgZn}.
In all these systems, (majority) Al/Zn plays the role of the ``small'' atom,
Mg is the ``large'' atom, and rare earths (RE) are even larger.
\MEMO{Mention the highly ordered icosahedral\cite{TNIMNTT94} 
quasicrystals in ZnMgRE here? (CLH put it in the conclusion.)}
Some icosahedral Frank--Kasper quasicrystals are well
understood, thanks to the solution of large ``approximant''
crystals (meaning the unit cell is an exact fragment of the
quasicrystal structure, and the lattice parameters have particular 
relationship to the quasilattice parameters of the quasicrystal).
Decagonal FK quasicrystals are less known;
the {\it stable} ones~\cite{Ni94,Tsai94b,A98,SAT98,AT99}
are the (presumably isostructural) $d-$ZnMgDy and $d-$ZnMgY;
the former has composition Zn$_{60}$Mg$_{38}$Dy$_2$.
\SAVE{(More exactly, 
Zn 59.8(4)\%,  Mg 38.5(4)\%, and Dy 1.7(2)\% atomic)}.
In general the composition is ZnMgRE
(where RE=rare earth, specifically Y,Dy,Ho,Er).
The compositions is
Zn$_{58}$Mg$_{40}$Dy$_2$\cite{AT00} for $d$--phase.

The structure was not solved from diffraction, as it
was impossible (until recently) to grow large 
(quasi)crystals~\cite{steurer-grow};
one powder diffraction fit was carried out, using
an earlier version of the structure model proposed in this paper
~\cite{suck-meltspun}.
Very recently, preliminary experimental results have been 
obtained on single (quasi)crystals of decagonal MgZnY~\cite{steurer-solve}.
Till recently, the most detailed structural information on $d$-MgZnY 
\MEMO{don't I mean d-MgZnDy?}
was Abe's~\cite{AT99} high-resolution electron microscope 
(HREM) images
\SAVE{(showing, in particular, the large-scale patterns of
pentagonal/decagonal clusters).}
Finally, some crystalline ``approximants'' of the decagonal are known
These include the structure types 
AlMg$_4$Zn$_{11}$, Al$_3$Zr$_4$,  Al$_2$CuLi, and Mg$_{4}$Zn$_{7}$.
\SAVE{
The last of these already has 110 atoms/unit cell.
(These approximants were not identified as such
in the literature.)}

In the present work,  we aim to {\it predict} the
detailed structure of the decagonal phase,
ultimately to the point that total energy
calculations can confirm the stability of the decagonal
with respect to other phases.  (Such comparisons are
notoriously difficult in quasicrystals, since a wrongly
occupied (yet rare) site may have negligible effects on 
diffraction and HREM, yet greatly increases the energy.)
Since $d$-ZnMgRE contains only 2\% of the RE, we initially
simplify the problem to binary Zn--Mg models. 
(The binary decagonal is already a plausible
phase, since it is {\it almost} stable 
with respect to competing ZnMg phases.)

We have three starting points:
\begin{itemize}
\item[(1)]
the Roth-Henley model~\cite{RH97}, which has a well-understood
and close similarity to the (better-understood) icosahedral 
structure 
\item[(2)]
two known, simple alloy structures are
in fact approximants of the decagonal structure.
\item[(3)]
electron microscope images give hints of the patterns
formed by repeated motifs, namely rectangle-triangle tilings.
\end{itemize}
\MEMO{The above list may be redundant with the same
list in Sec.~\ref{sec:decoration-atoms}.}

\SAVE{Ref.~\cite{RH97} introduced a hypothetical FK decagonal
model, idealized from molecular dynamics simulations 
of a toy binary alloy.  No experimental realization
was known at the time, but is subsequently
turned out to capture the gross features 
of the real FK decagonals.}

Since the structure is decagonal, the structure is 
quasi-two-dimensional (the lattice parameters from electron
microscopy show the period is two atomic layers).
We adopt the {\it tile-decoration} approach to specifying
structure models.   This means all space is tiled by copies 
of two (or a few) polygons (the ``tiling''),
and in turn each kind of polygon 
contains a set of sites on which atomic species are assigned,
analogous to the Wyckoff sites of an ordinary crystal
(the ``decoration'').
This approach has the following advantages:
\begin{itemize}
\item[(i)]
only a limited number of real parameters are needed to
specify the atom locations 
\item[(ii)]
instead of attacking the 
structure problem in one step, we divide it into two parts
which can be addressed independently 
\item[(iii)]
by using various
(periodic) tilings coupled with the same decoration, we can 
construct a large family of ``approximant'' crystals with
closely related structures. Some of them are realized in
nature, and all of them are convenient for numerical 
calculations of the total energy (or other properties),
which are difficult to set up for an infinite, aperiodic system.
\end{itemize}
The old structure model~\cite{RH97} was also of the decoration
type, but was based on the Penrose or HBS (Hexagon-Boat-Star) tilings, 
whereas we now find the natural tiling for $d$-ZnMg
is a rectangle-triangle (RT) tiling.


Throughout, we will use ab-initio calculations with the 
VASP package~\cite{vasp} as the basis for our fitted pair potentials, 
and also use ab-initio
energies for computing phase stability of the final structures.

\section{Basic atomic structure as  tile decoration}
\label{sec:tile-decoration}

The basis of our structure model is that all the low
energy structures are tilings of the plane by two kinds of
tile, the Triangles (T) and Rectangles (R) shown in Figure~\ref{fig:deco}.
The main tile edges (heavy lines in Fig.\ref{fig:deco})
have length (on the plane) 
$\cong 4.5$xx\AA, and are oriented at angles differing by
multiples of $2\pi/10$. 
A second kind of edge (dashed lines in the figure).  
of length $b\equiv 2 a \sin(2\pi/10) \cong$ 5.2\AA~
points along diections halfway between the ``$a$'' edge directions.
End each kind of tile always has the same ``decoration''by atoms.  
The three-dimensional structure is formed by repeating the pattern 
perpendicular to the plane with period $c=...$\AA; each period contains 
the equivalent of two atomic layers.

\subsection{Decoration rules}
\label{sec:decoration-atoms}

\SAVE{(From MM 4/27/11):
To a first approximation, the structure of all FK quasicrystals (also dodecagonal)
is controlled by nearest-neighbor interactions, and can be roughly described in terms
of sphere packings. 
(which would be Monodisperse for dodecagonal cases, or  binary large/small for decagonal.
But CLH notes, the sphere packing should be non-additive hard spheres.}

Each tile vertex has a column of small atoms, two per lattice constant (ZnZn),
with $z$=$\pm$0.25. [Coordinates $z$ are given in multiples of 
the stacking periodicity $c=5.1$\AA.]
Each mid-edge also has a Zn at $z$=0 or 0.5 (we will explain shortly
how this is decided).
Finally, the tile interiors have sites which are all large (Mg) atoms,
also at $z$=0.5 or 0).  

The recipe for the $z$ heights is simple. 
Consider all the network of non-vertex atoms: as seen
in projection, it consists of one irregular decagon 
around each vertex, plus one irregular square in the
middle of each Rectangle tile.  Both polygons have
an even number of vertices, so the network is 
bipartite; we simply assign height 0 to the even
vertices and height 0.5 to the odd vertices.
This ensures that these atoms form pentagonal
antiprisms around all the Zn on vertices.
\SAVE{(Hence the latter have coordination CN=12.)}
A corollary is that, on either side of an $a$ edge, we have Mg atoms 
at the {\it same} height, different from that  of the mid-edge Zn between them;
on either side of a $b$ edge, we have Mg atoms at different heights.

The rule for the $z$ heights can be restated as a {\em global} pattern:
the mid-edge Zn has height $z=0$ when the orientation angle of
its $a$ edge is an {\it even} multiple of 2$\pi$/10 (as viewed from
an even vertex towards an odd vertex) or $z=0.5$ when the angle
is an {\it odd} multiple of 2$\pi$/10 angle.

\begin{figure} 
\begin{center}
\includegraphics[width=2.1in]{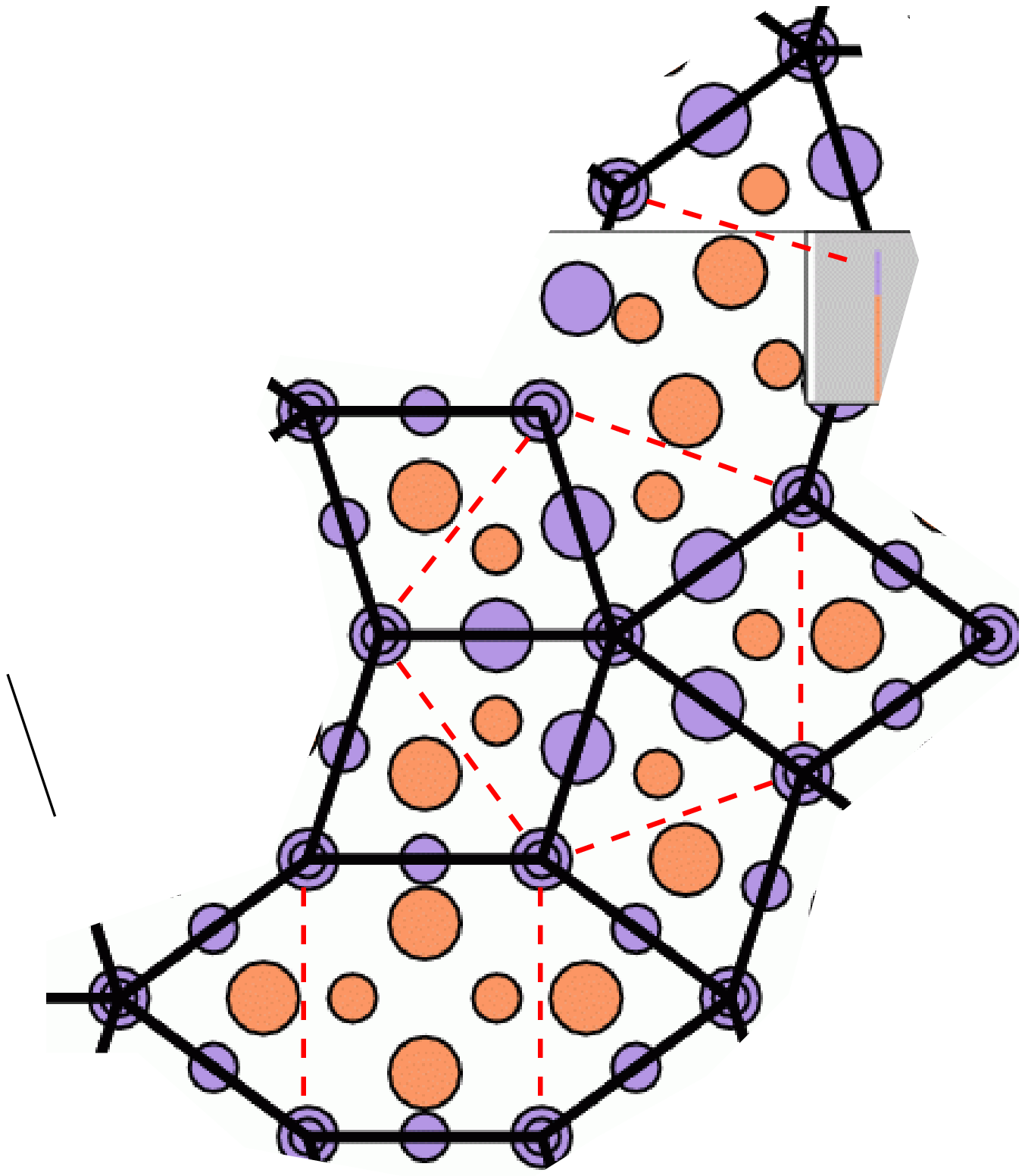}
\caption{
\label{fig:deco}
Atomic decoration of the \FrH\ tiles. 
Zn atoms are shaded darker.
Larger circles are {\it lower} in height along (pseudo)decagonal axis. 
Solid (dashed) lines are $a$ ($b$) type linkages respectively. 
The rectangles combine into Hexagon and Fat Rhombus tiles (solid edges);
in turn, five Fat Rhombus tiles form a Star. 
\MEMO{Figure is a placeholder.  The intent is to rotate
it horizontal, also add a key "Mg .../Zn ...".
Notice that Rectangles are shown with both possible levels.
CLH might make the Mg a little lighter to have a clear contrast
in case of black-and-white.}
}
\end{center}
\end{figure}


There are three inputs suggesting this model:
\begin{itemize}
\item[(1)]
The Roth-Henley decagonal model\cite{RH97} was derived from a
Monte Carlo simulation of a large and small species of toy atoms 
interacting by Lennard-Jones potentials (only a nearest neighbor
well), with well radii chosen so as to favor Frank-Kasper structures.
That paper emphasized a version based on the Hexagon, Boat
and Star tiles, which are combinations of the Penrose rhombi.
As Ref.~\cite{RH97} already noted, this is already practically
a Rectangle-Triangle tiling:
as seen in Fig.~\ref{fig:deco}, the Hexagon tile breaks up into
a Rectangle plus two Triangles, and the Fat Rhombus breaks up
into two Triangles.  (The Star consists of five Fat Rhombi sharing
their tips.)  The main difference from our present simpler model
is that Ref.~\cite{RH97} also used a Boat tile (which is further
discussed in Appendix~\ref{app:alternative-deco}), below.
\item[(2)]
Several known crystalline phases have a packing of Rectangles and
Triangles as their unit cell.
In particular, all Laves phases correspond to pure-triangle tilings
(in particular the stable Laves phase MgZn$_2$.)
\item[(3)]
Our own examination of the high-resolution electron microscopy
(HREM) images from Ref.~\cite{Abe-HREM} found them to be largely
made of Rectangles and Triangles.  (Of course, this third input
does not indicate the atomic decoration rule, except that is
is known the bright dots correspond to the columns with two Zn
atoms per period.)
\end{itemize}

Most of the known decagonal approximants are triangle-hexagon tilings (hex.=rectangle+2tri.).
Examples of large decagonal approximants.

\begin{table}
\caption{Stoichiometry of the tilings. $N_{tri}$ and i$N_{rec}$ 
are numbers of triangles and rectangles respectively, 
N is number of nodes in the unit cell, $x$ 
fractional content of the species in the alloy. In the last row, 
$tau$=(1+$\sqrt{5}$)/2 \label{tab:summary}}
\begin{tabular}{cccccc}
tiling        & N  &  $n_{tri}$/N & $n_{rec}$/N & $x_{\text{Zn}}$ & $x_{\text{Mg}}$\\
\hline
triangle    & 1  &  1    & --    & 2/3 & 1/3 \\
rectangle         & 2  & --    & 1/2   & 3/7 & 4/7\\
\mgivznvii\    & 18 & 1.778 & 0.111 & 0.636 & 0.364\\ 
$\tau$-ZnMgDy  & 50 & 1.680 & 0.160 & 0.623 & 0.377\\ 
$d$-ZnMgDy     & 1 & 4$\tau^{-2}$=1.528& $\tau^{-3}$=0.236 & 0.603 & 0.396\\
\end{tabular}
\end{table}

\MEMO{Is this true?}
The validity of the basic decoration has also been 
confirmed by ``melt-quench'' MD simulations~\cite{MQ-AlCoNi,EOPP}
in which the system 
is simply cooled from high temperatures, in a cell of properly
chosen dimensions so that an  ordered (quasicrystal) structure is 
obtained rather than a glass.

\MEMO{``The decagonal FK structure is strikingly different from Al--based 
quasicrystals [...] in that its structure is NOT robustly impacted
by formation of large clusters''.  Firstly, do you mean Al--TM? 
But I didn't think those decagonals have large clusters, necessarily.
I thought the dichotomy large cluster/smaller cluster goes with
ico/decagonal.}

\MEMO{from MM 4/27/11}
``Frank-Kasper quasicrystals show almost no occupational disorder 
(in glaring contrast with the Al-transition metal kind).''
\MEMO{to MM: you wrote Al-based, did you mean to contast ZnMg with 
AlMgZn?  It seems to me this is mainly a consequence of being a binary
rather than a ternary with similar size atoms;  also, not having
abundant pseudo-vacancies as found in Al-TM.}

\SAVE{The Laves phases are common grounds between icosahedal and decagonal FK.}

\SAVE{The Roth-Henley decoration, in turn,
was inferred from the simulation of a toy atomic model~\cite{RH97}.
It is also a slight deformation of a
structure model~\cite{HE86}
that was well known to capture the elementary efeatures of 
{\it isosahedral} Frank-Kasper quasicrystals.}

\subsection{The rectangle-triangle tiling}

\begin{figure} 
\caption{
\label{fig:known-tilings}
Known Zn-Mg structures related to rectangle-triangle tilings.
(a) \mgivznvii\  (b) $\tau$--ZnMgDy and
(c) decagonal ZnMgDy; the latter two are reconstructed from HREM
images from Ref. CITE}
\begin{center}
    \includegraphics[width=3.2in]{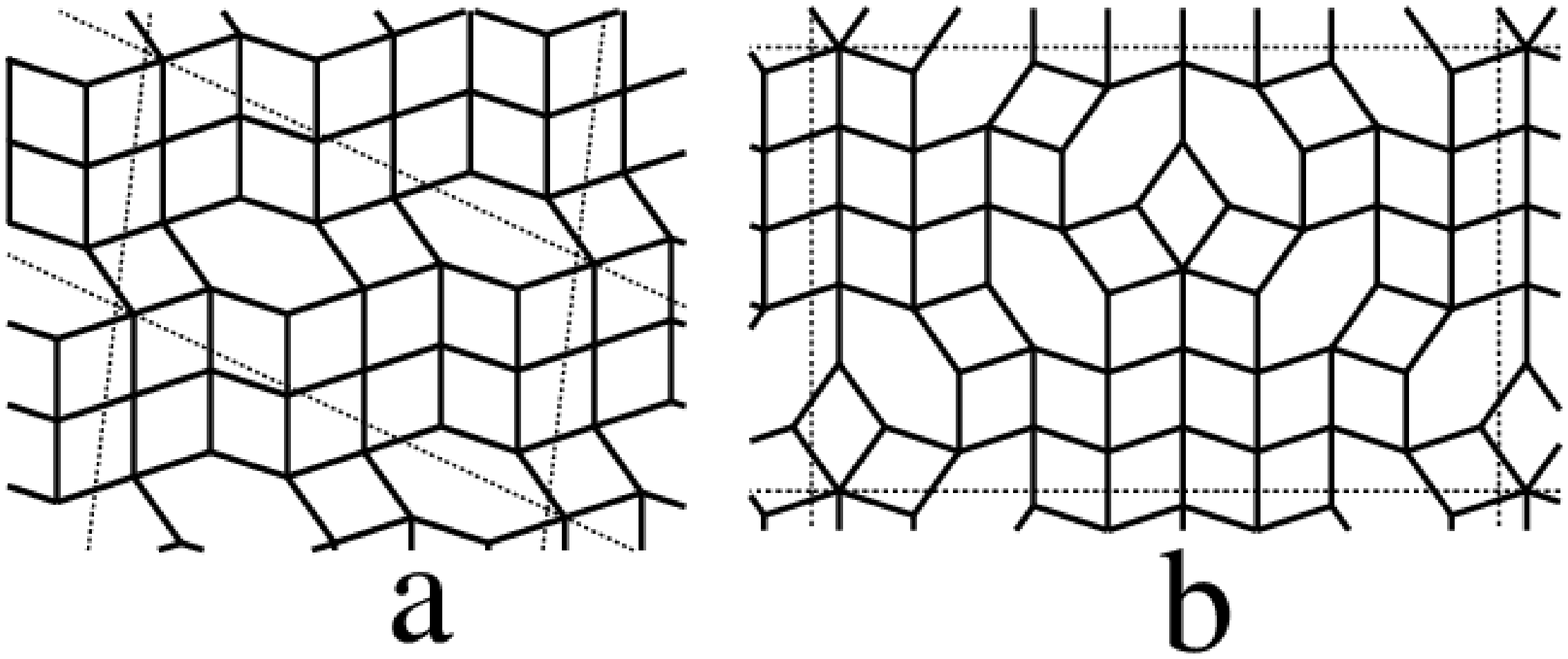}
    \includegraphics[width=1.6in]{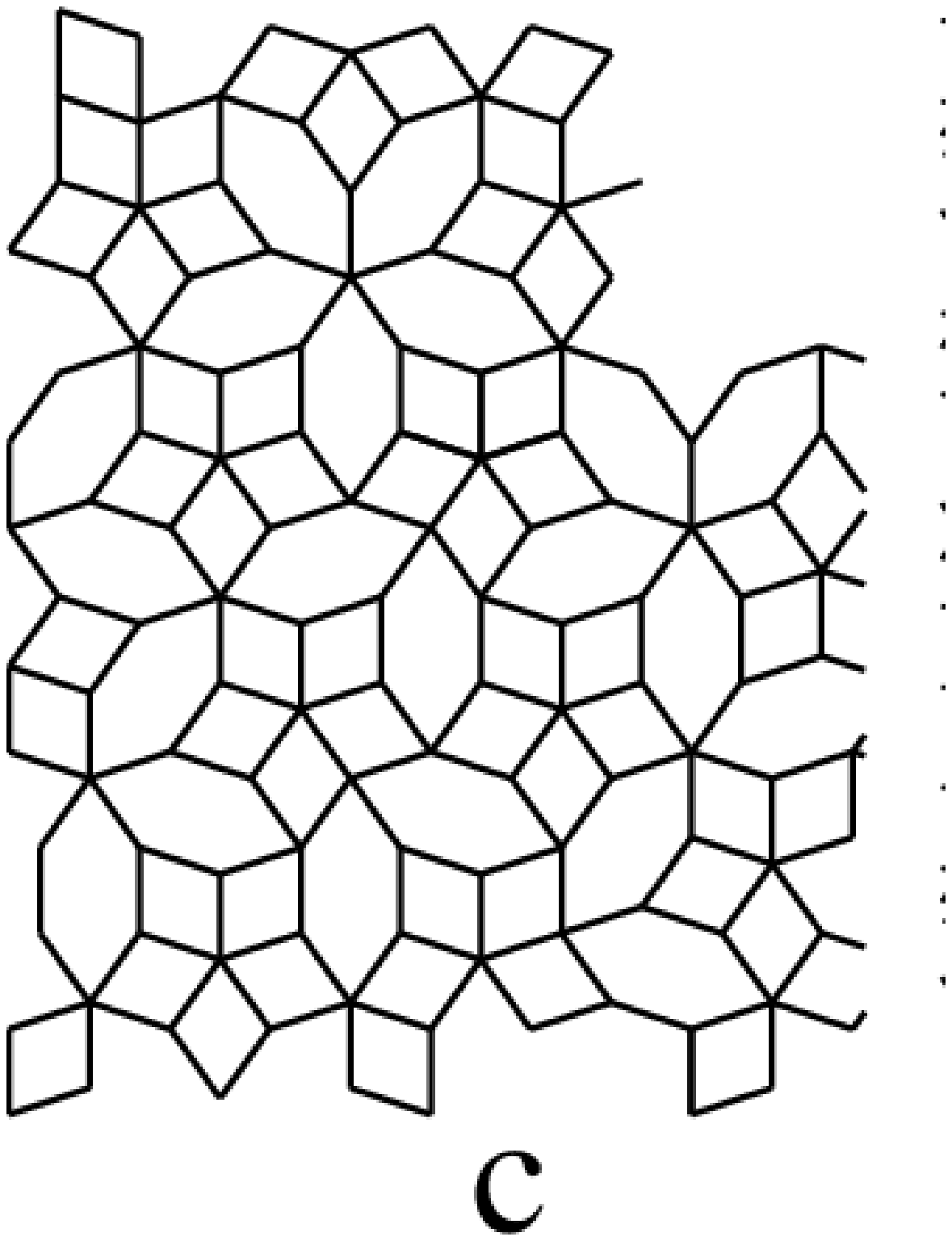}
\end{center}
\end{figure}

Given these decoration rules, the R--T
tilings are in one-to-one 
correspondence with a class of atomic structures
including approximants and quasicrystals. The crystalline \mgivznvii\
has a structure exactly corresponding to the tiling in 
Fig.\ref{fig:known-tilings}(a).
The decagonal  phase forms upon annealing of metastable ternary 
\mgivznvii\ structure, and the $\tau$-approximant is a 
transitional state \cite{AT00}.
Fig.\ref{fig:known-tilings}(b) and (c) 
are reconstructed from the HREM images of 
Refs.~\cite{AT00,AST99}, 
in which the authors justified the correspondence
between the images and the atomic structure by simulating images
at an appropriate defocus condition.
\MEMO{CLH needs to look up}

Table \ref{tab:summary} summarizes the stoichiometry of the tilings
assuming a purely binary ZnMg decoration rule. 

The Mg atoms next to the $a$ edge in the R tile is noticeably
squeezed. Indeed, this site has coordination CN14, which is 
typically on the borderline between small-atom and large-atom
sites in a Frank-Kasper structure (the analogous sites in 
icosahedral-related structures have large atoms in 
the AlZnMg system but small atoms in the AlCuLi system).
This disfavors environments in which the other side of
the $a$ edge is also squeezed.

\MEMO{MM wrote:``A possible explanation of 
why larger approximants or quasicrystal do not form in binary system.''
Are you saying that large approximants demand R--R pairs?  
I don't think that is true.}

\section{Fitting the tile Hamiltonian}

It is feasible to obtain ab-initio relaxed energies of the decoration
structures  from relatively small R-T tilings, but not for an exhaustive 
enumeration of them.  Hence we pursued an effective Hamiltonian approach,
starting with a database of the ab-initio energies for 155 R-T tilings 
(decorated and relaxed).  These  energies were fitted to an effective
Hamiltonian $\HHtile$, written as a sum of local interaction terms depending only
on the tile degrees of freedom.

\subsection{Ab-initio based fitted pair potentials}

\begin{figure}
\includegraphics[width=2.7in]{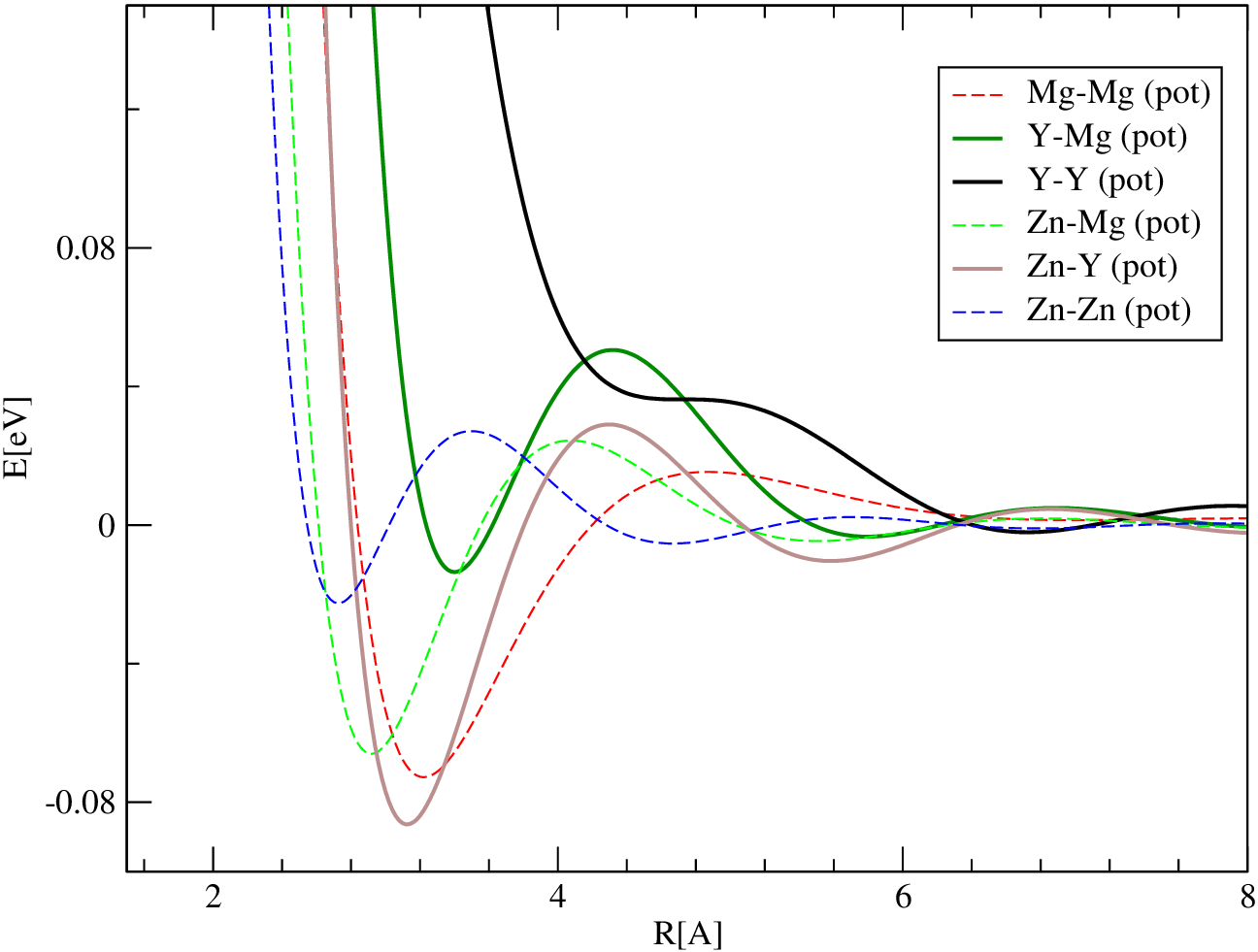}
\vskip 0.2truein
\caption{
(a) Fitted pair potentials for binary Mg--Zn.
Notice, around 4\AA~(??), the significant second well in the 
Zn-Zn potential and hump in the Mg-Mg potential [as well as the
strong Zn-Mg attraction, but where do we use that?]
(b) Additional potentials involving Y (other rare earths
such as Dy are similar).
\MEMO{Place-holder figure.  The actual figure shows 
all six potentials on one plot.  The figure also has
white space beyond its bounding box so as to cover up
the caption.}
}
\label{fig:potentials}
\end{figure}

\MEMO{A text sectioning problem.  Should this be a
separate section on methods? It doesn't quite belong
with the tile Hamiltonian, but seems so
short that I made it a subsection -- what do you think?
(Basic reference to VASP is made in the introduction.
CLH deferred the ``zipper'' method
until Sec.~\ref{sec:tiling-results} where we use it.)}

``Empirical oscillating'' pair potentials (EOPP)~\cite{EOPP}, 
were fitted to a database of ab-initio forces (at $T>0$)
and energies (for $T=0$ relaxed structures) calculated using VASP~\cite{vasp}.
\MEMO{More about the database -- how many samples, what composition range???}
This was extended to ternaries with Y; Figure~\ref{fig:potentials} shows the 
results.  
Since both Mg and Zn have effective valence $+2$,
the electron/atom count is indepenent of composition, 
so the potentials should be valid across a wide composition range.
\MEMO{Needed? ``perhaps that is why the Mg-Mg and Zn-Zn potentials are remarkably 
close to those for the pure elements.''}
(Some other FK quasicrystals, in which the majority constituents 
are simple $sp$ metals, are even better adapted to 
pair potentials~\cite{hafner}, as was tried long ago but
without attempting  any fine details of atomic structure.)


The potentials have minima at 3.3\AA~ [for $V\MgMg(r)$] or 
xxx for $V\MgZn(r)$], while $V\ZnZn(r)$ has only a 
shoulder at the nearest-neighbor distance.
(Typical actual distances are 3.0--3.3\AA~ for Mg--Mg, 
3.0\AA~ for Mg--Zn, or 2.6\AA~ for Zn--Zn.)
The Zn--Zn potential has a a strong second-neighbor well at 4.5\AA,
while the other two potentials have humps at $\sim 1.25$ times
the nearest-neighbor distance. 
\SAVE{A hump around $\sqrt 2$ of that
distance discourages structures with 90$^\circ$ angles,
such as fcc or bcc [cite Dzugutov?]}

Till now, it has been unclear whether the oscillations 
in the pair potentials are essential for quasicrystal order 
in the FK class of alloys.
A positive hint is that both the Al-TM and FK classes 
are believed to be stabilized by the ``Hume-Rothery'' 
mechanism~\cite{friedel-FK}:
that means that (within the nearly free electron picture)
the ions arrange themselves so that the potential 
seen by electrons has prominent scattering 
at wavevectors that span the Fermi surface, thereby
mixing electronic states near the Fermi energy so
as to create a pseudogap there and lower the energy
of the occupied states.   
Expressed in terms of the real-space structure, this 
is equivalent to saying the pair potentials have strong
(``Friedel'') oscillations and that 
their second or third wells play a role 
in selecting the structure.
Indeed, the second potential wells were shown to be
crucial in the Al-TM case~\cite{Lim08a,Lim08b}.

\SAVE{A negative hint is that simulations using
Lennard-Jones potentials, which have only the
first-neighbor wells, did reach approximations of icosahedral and decagonal
phases~\cite{Roth95,RH97}, however that could be a kinetic
selection effect; it is not known whether such phases 
are ever thermodynamically stable according to such potentials.}

\SAVE{Presumably the pair potentials can be used to show
why the basic (decorated) rectangle-triangle structure is
good at all, but we do not attempt that here: we take the 
R--T tiling and decoration as a given.}

\begin{figure}
\caption{The ten allowed vertex types in the rectangle-triangle tiling, 
$V1$,...,$V10$.
The ``$a$'' edges are shown by straight lines (red online)
and the ``$b$'' edges by double lines.
The standard names shown are strings of four to six numbers $\alpha_i$,
such that $\alpha_i(\pi/10)$ is the angle at corner $i$, in order around
the vertex.  One example is labeled for each of the seven edge types,
e.g. $a_{RT}$ ($a$ edge shared between a rectangle and a triangle).
Note the distinction of $a_{TT}^m$ (mirror edge) and $a_{TT}^z$
(``zigzag'' edge, inversion center at its midpoint).
Boxes are drawn around the five vertex types appearing in optimal tilings.
\MEMO{Probably we should have a consistent graphical language
in all figures for the $a$ edge and $b$ edge.}
}
\label{fig:vertices}
\begin{center}
   \includegraphics[width=4.4in]{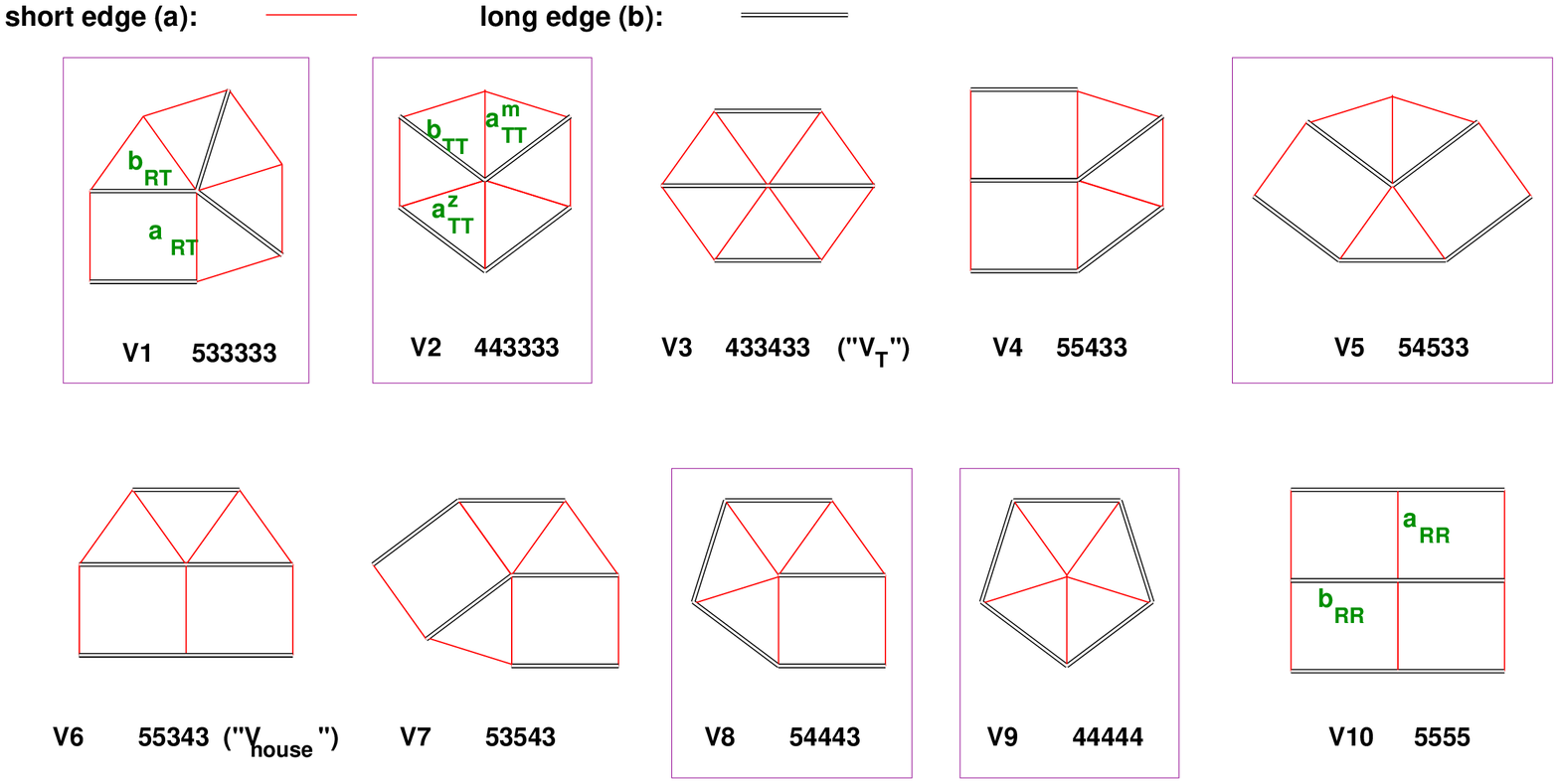}
\end{center}
\end{figure}

\subsection{Tile Hamiltonian form and fits}

The tile-Hamiltonian approach\cite{mihet96a} posits that the 
stronger interatomic interactions favour the atoms to organize 
into a decorated tiling  (in our case the \rt\ tiling) which 
are practically degenerate.  The terms of the Hamiltonian 
that distinguish different tilings are weak enough to
be treated as a perturbation.  We can collect these contributions
and attribute them to the tiling geometry, thus defining
a ``tile Hamiltonian'' acting on these reduced degrees of 
freedom (which correspond one-to-one to the possible low
energy atomic configurations).  The tile Hamiltonian's
low-energy states will define a ``low-temperature'' subensemble,
as analyzed below in Sec.~\ref{sec:tiling-results},
which might be built from ``super-tiles'' which are clusters
of the basic tiles.

We consider three possible kinds of terms in $\HHtile$:
\begin{itemize}
\item[(a)] 
Single-tile terms,  $E_R N_R+E_T N_T$, where
$N_R$ ($N_T$) is the total number of rectangles (triangles).
Note that, with the decoration we are using, there is a linear
relationship between $(N_R,N_T)$ and the atom counts
$(N\Mg,N\Zn)$, so this term is equivalent to chemical potentials
coupling to the atom numbers.
\item[(b)] 
Tile-pair ``edge'' terms, such as $E(b_{RR}) N(b_{RR})$,
for each of the seven possible edge types shown in
Figure~\ref{fig:vertices}.  Note there are four linear
dependences among the counts of these edges and $(N_R,N_T)$.
An example is $2 N_R = 2N(a_{RR})+  N(a_{RT})$, which simply
expresses the fact that each rectangle has two $a$ edges, and
each of them is shared with either another rectangle or a triangle.
\item[(c)]
``Vertex'' terms, such as $E(433433) N(433433)$, where
$N(433433)$ is the number of one of the ten vertex types
shown in Figure~\ref{fig:vertices}. This incorporates
not only second-neighbor tile interactions between tiles that
share a corner, but also multi-tile interactions among
all the tiles sharing that vertex.  There are additional
linear dependences among the counts of these vertices,
most strikingly $N(54533)=N(44444)$. 
(See Appendix~\ref{app:sum-rules}).
\end{itemize}

\SAVE{This follows because every edge in any vertex can 
be unambiguously assigned to an edge type; no other information
about the tiling is needed.}

In fact, any tile-pair term can 
be rewritten entirely in terms of vertices.
\SAVE{using the recipe that each vertex gets half the edge energy 
from each of the edges that meet at it.}
Though our actual fit was to the vertex terms, we re-expressed
the results (as much as possible) in terms of
single-tile terms and edge terms.
In resolving the many linear dependencies between counts,
we chose a parametrization that made the coefficients positive and
small.   At the end, the largest terms were the single-tile energies,
and edge terms for $a_{RR}$, $b_{RR}$, and $a_{RT}^z$;
There are also two smaller ``vertex'' terms tending to disfavor vertices 
$\langle 433433\rangle$ and $\langle 55343 \rangle$, 
which we shall call $V_T$ and $V_\house$ for short.

For the fit, we used the pair-potential energies for
three groups of tilings, having (a) 12 tilings of 29 vertices, 
(b) 63 tilings of 44 vertices
\SAVE{The 44-vertex tilings were in $2\times 2$ enlargements of the 
unit cell of the unique 11-vertex approximant.}
and (c) 80 tilings with 47 vertices.  

The fit results are shown in Table~\ref{tab:Htile}.
(The uncertainties are the variance between the fitted parameters 
using any two of the three groups.)
\MEMO{to Justin: Need to calculate uncertainties properly
and put them in the table.  The uncertainties in present text
are actually 2/3 of the difference of the even/odd vertex number
tilings.}

\begin{table}
\caption{Fitted energies for terms in tile Hamiltonian (eV)}
\label{tab:Htile}
\begin{tabular}{|l|rrrrrrr|}
\hline
Hamiltonian & 
          R  &  T &  $a_{RR}$ & $b_{RR}$ & $a_{TT}^z$ & $V_T$  & $V_\house$ \\
\hline
Vertex  & $-0.899$ & $-0.620$ & 0.134 & 0.102 & 0.030(?) & 0.014 & 0.018\\
\hline
\end{tabular}
\end{table}

\subsection{Explanation of tile Hamiltonian results: second potential well?}

The size and sign of the tile Hamiltonian terms can be understood
by comparing the pair potentials to the interatomic distances.
The largest terms, by far, are disfavoring either kind of 
Rectangle-Rectangle edge, with $a_{RR}$ being twice as bad as $b_{RR}$.
\SAVE{The former edge would be allowed as a degree of freedom in the 
Hexagon/FatRhombus version of this tiling, the latter would not be.}
This is because the large interior (Mg) atoms in the Rectangle are 
too tightly packed; this is accomodated by outward displacements when
the adjoin Triangle tiles, but not when they adjoin other Rectangles.

\MEMO{So far, the conclusions of the next paragraph are unclear.}
Are interactions significant beyond the nearest neighbors?
To test this, we re-ran the fits using pair potentials cut
off at 4.0\AA (between the first and second neighbor distances),
using the relaxed energies.  This had negligible effects on the
cost of $a_{RR}$ or $a_{TT}^z$ 
but greatly improved that of $b_{RR}$ (from $0.081 \to 0.013$ eV).  
\SAVE{The $N_T$ term changes a bit ($-0.538 \to -0.575$ eV), but cutting
off the potentials greatly improved the $N_R$ energy ($-0.980 \to -1.398$ eV).
As stated earlier, this is equivalent to adjusting the chemical potential
for the two atomic species.}

\MEMO{Go on to explain the tests Justin deployed, and our
specific interpretations how the zigzag term in the tile Hamiltonian
comes from the second well.}

We can clearly see the explanation of the ``zigzag'' energy cost.
Note first that, in the idealized pattern, the Mg atom in either
triangle is exactly lined up with the center of the edge, so its
position is unchanged when one of the triangles is reversed 
(so as to convert the ``zigzag'' edge to a ``mirror'' edge).
The nearest atoms that do change their relative position are
the mid-edge Zn atoms on the far sides of each Triangle.
In the ``mirror'' configuration, these Zn sites are separated 
by a distance $(\tau/2)b\approx 4.25$\AA, which is close to the
second Zn--Zn potential well; in the ``zigzag'' configuration,
they are separated by $a\approx 4.5$\AA, which 
is close to a hump in the Zn--Zn pair potential.
\MEMO{Check those numbers}.
This shows that the ``zigzag'' term -- the key interaction
giving the organization that we find -- indeed depends on
the oscillations in the potentials.

\subsection{Placement of rare earth Y in the ternary}
\label{sec-whereY}

The experimental quasicrystal phase (CITE) is
stabilized by a just a few percent of Y atoms,
which are known to be effectively even larger than Mg,
and (we expect) are filling special favorable sites.
\SAVE{(The rare earths are indeed rate in this phase!)}
fill particularly advantageous sites in the structure.
Thus our key choice in devising the Mg--Zn--RE 
decoration rule is, which out of the CN=16 Mg sites
are to receive the rare-earth atoms? 

We found that the most favorable site to substitute
Mg $\to$ Y is in a triangle which forms the ``4''
corner of the $\la 54533 \ra$ vertex.  
\SAVE{(Thus there is one such site per $\la 54533 \ra$ vertex.)}
It can be shown that, in fivefold symmetry, such sites
are a fraction $0.5\tau^{-5}(\tau+1/4)\approx 8.42\%$ 
of all Mg sites, i.e. the net Y content is 3.4\% atomic,
which is slightly larger than the experimental content.
\MEMO{Is my percent correct?}

We can understand this placement of the Y atom in the
following fashion.  Firstly, the Mg sites in the Rectangle
are overpacked (indeed these have CN=15) and Y is bigger 
than Mg, so the favorable site is Mg in the Triangle
(which has CN=16).  Secondly, the Y--Mg potential
(Figure~\ref{fig:potentials}) is quite repulsive at 4.5\AA,
and the Y--Zn potential is repulsive at 4.3\AA.  Say the
neighboring tile by the $a$ edge is another triangle.
If that is part of a Fat Rhombus, there is another Mg
site (at the same $z$ height) in the far triangle of the
Fat Rhombus, at a distance from the first site of
ideally $a\approx 4.5$\AA which is disfavorable.
(If the adjacent triangle were part of a Hexagon,
that would still have an Mg site in almost the same place.)
\SAVE{In addition, there is a mid-edge site on the adjacent
triangle shifted in $z$ by $c/2$ and a distance
$0.5 \sqrt{1+3 \tau^{-2}}a$ in the plane,
so the total distance is about $4.8\AA, which is 
somewhat disfavorable.}
\MEMO{I should check what interatomic distances are
found on the other side of the $b$ bond.  In this
site, that is invariably another triangle, being
part of a Fat Rhombus: check why being in a Hexagon
would be bad.}
Thus it is favorable for the triangle with Y to be adjacent 
to a Rectangle: in the triangle adjoining an $\la 54533 \ra$,
it is next to {\it two} Rectangles (the maximum).

\MEMO{Does this modify parameters in the tile Hamiltonian?
By how much?}

\SAVE{An obsolete speculation was that a cluster of
small Zn atoms could replaced by one RE atom, but we
did not find any favorable place to do this.
Such an exercise would be an extension of the 
investigations in Appendix~\ref{app:alternative-deco}.}

\MEMO{(MAREK: what is your comment on this old speculation?)
The chemical composition of the ternary phase 
was Zn$_{58}$Mg$_{40}$Dy$_2$\cite{AT00}; since the
binary decoration model also gave $x_{\rm Mg}=0.40$,
(see Table \ref{tab:summary}), this suggested
an effective Zn$\rightarrow$Dy substitution.
We speculated that Dy alleviates the packing problems of the CN=14
atoms in Rectangles, or that Dy allows the presence of
combination tiles that contain the equivalent of an unpaired Skinny rhombus.}

\section{Minimum energy tilings}
\label{sec:tiling-results}

Having obtained the tile Hamiltonian, the next task is
to discover (and characterize) which tilings minimize it,
by Monte Carlo simulations in medium and large cells.
\SAVE{(For technical reasons, having to do with our representation
of the tilings, the Hamiltonian was converted to the
pure vertex form in all cases.)}
This is far more challenging technically than for
tilings of Penrose rhombi, or of Hexagon/Boat/Star tiles,
in which any tiling can be accessed from another
via a string of {\it local} reshufflings involving
two or three tiles.  In the case of the R--T tiling, 
a nonlocal move called a ``zipper'' is necessary, 
encompassing a chain two tiles wide (and typically 
recrossing itself) that continues until the chain reaches 
the starting point and closes on itself.~\cite{OxHen93,OM95,OM98}.
The implementation for the R--T case~\cite{OM95,OM98} is
similar to that for the 12-fold symmetric 
square-triangle tiling~\cite{OxHen93}, but more complex.

\MEMO{CLH: do we want to show a small figure illustrating
a zipper move?}

\subsection{Results}

\begin{figure} 
\caption{
\label{fig:MC-tilings}
(a) Low temperature tiling of the 76--node approximant
obtained by Monte--Carlo annealing.
Only $a$ edges are shown in these plots.
(b) Supertiling.  Its nodes are centers of Stars
of five Fat Rhombi; even and odd nodes are 
marked by closed or open circles.
Four kinds of edges (right) are indicated by different
lines, and form three kinds of triangular tiles called
A, B, or C.
\MEMO{Semi placeholder figure.  Do we show more examples
(or different ones)?  Do we make supertiling a separte figure?
Note I changed the names of the edges, so that edge
$\beta$ goes in between B triangles and edge $\gamma$
goes in between $C$ triangles.}
}
\begin{center}
    \includegraphics[width=1.6in]{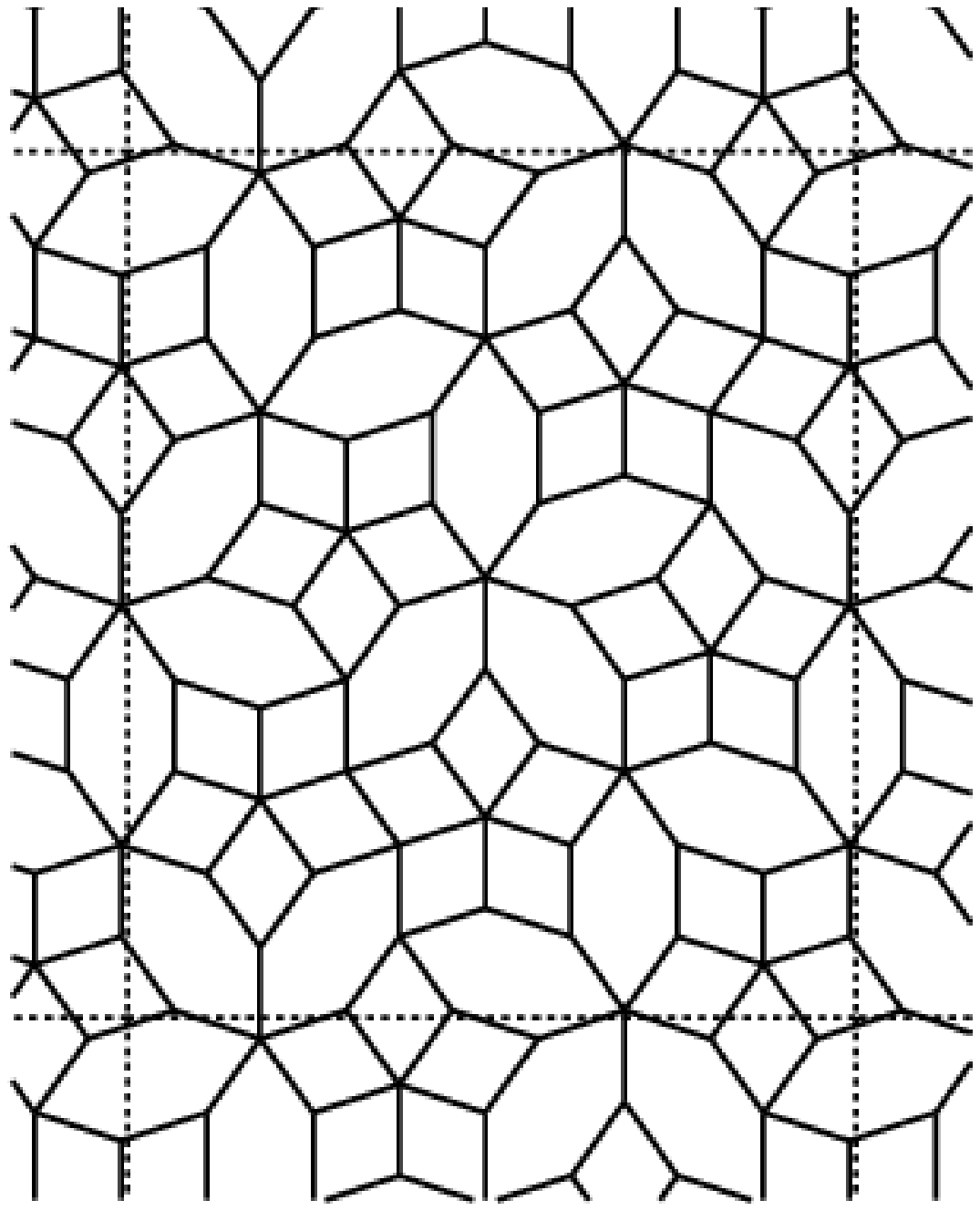}
    \includegraphics[width=2.6in]{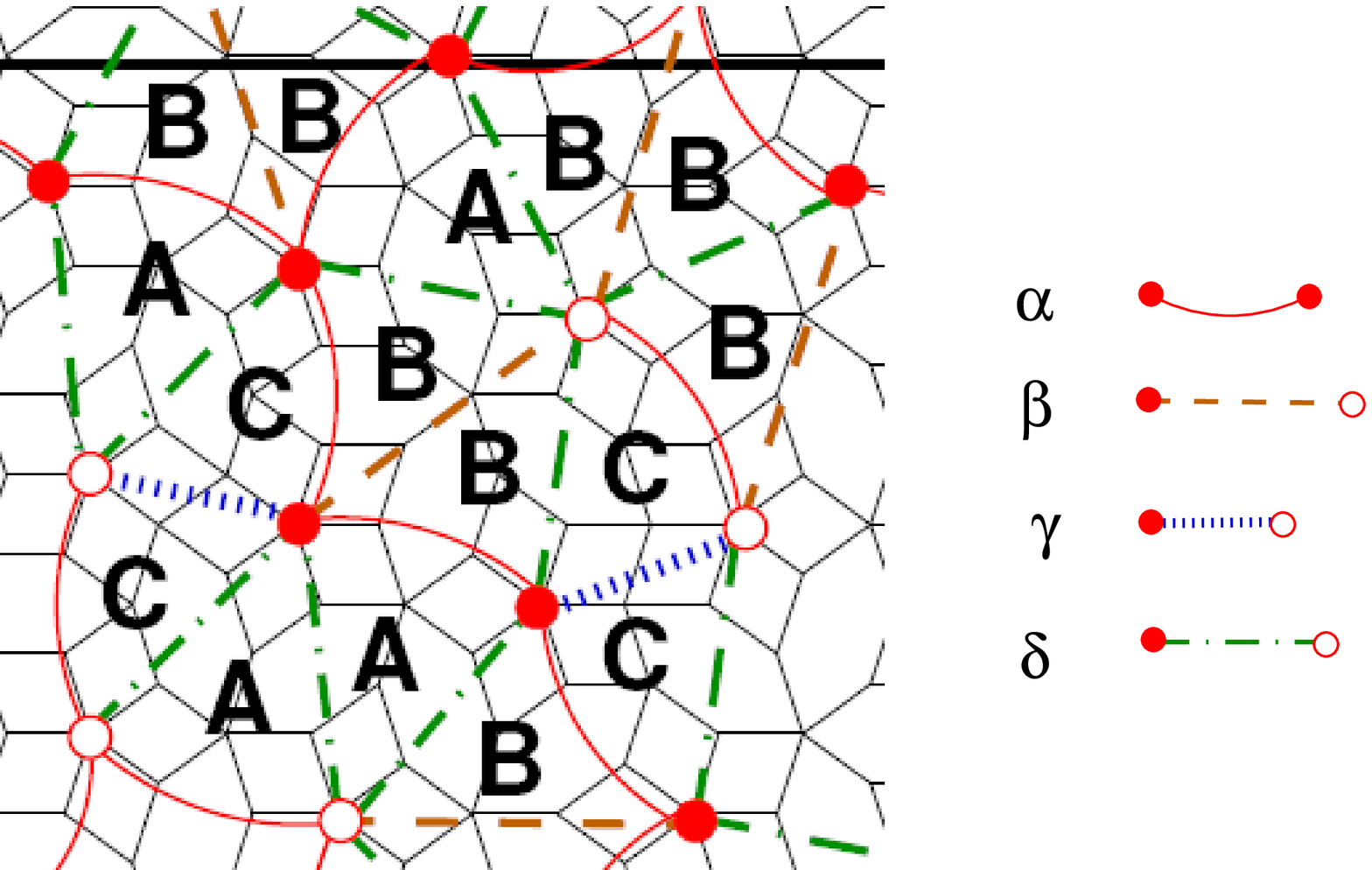}
\end{center}
\end{figure}

The resulting tilings show a great deal of local order
(see the example in Fig.~\ref{fig:MC-tilings}).
Some of the order can be understood immediately from the
tile Hamiltonian.  Most obviously, there are {\it no} rectangles
adjoining by either kind of edge, as expected since those
edges are very costly.  An immediate geometric consequence is
that the rectangles and triangles always combine into Fat Rhombi or
Hexagons (as in Figure~\ref{fig:deco}).
We also find that only five of
the ten vertex types appear, as
indicated by the boxes in Figure~\ref{fig:vertices};
three of these absent vertices are obvious, as they
included adjoining Rectangles.  

Given there are no R--R edges, the counts of most kinds of edge
are constrained through sum rules that follow from obvious
facts: e.g., every rectangle $b$ edge has a triangle on
the other side, and every triangle $b$ edge not adjoining
a rectangle must be of $b_{TT}$ type, adjoining another triangle.
(For details about sum rules see Appendix~\ref{app:sum-rules}).
However, the sum rules do not distinguish the two ways
that triangles can adjoin by an $a$ edge: with mirror
symmetry ($a^m_{\rm TT}$) or with twofold symmetry 
($a^z_{\rm TT}$); we call the latter ``zigzag'' from the
pattern formed with the opposite tile edges.
The largest remaining term in the tile Hamiltonian
(Table ~\ref{tab:Htile})
disfavors the zigzag edge, which probably explains the
absence of the $\la 433433 \ra$ vertex (which contains
four ``zigzags''). We have no explanation for the 
last absent vertex type $\la 53543\ra$.

\MEMO{Note that if two Fat Rhombi adjoin so as to be 
parallel -- a local pattern inimical to most kinds of
quasicrystal order -- they form a zigzag edge. Thus,
in disfavoring that edge the tile Hamiltonian {\it tends} 
to favor quasicrystal-like states with fivefold 
statistical symmetry and long range order, but
not in the absolute way that a matching rule does.}

Notice there only one term of the tile Hamiltonian -- the
zigzag term -- takes nonzero values in a tiling constrained
to use the five boxed vertex types from Fig.~\ref{fig:vertices}.
It follows that {\it our ensemble has no numerical parameters}
and is defined by minimizing (subject to the other constraints)
the number of zigzags (they cannot all be eliminated).

\SAVE{The same ensemble could be the
optimum for a variety of tile Hamiltonians. 
In particular, a made-up tile Hamiltonian with
terms disfavoring R--R edges and zigzags,
and favoring the fivefold vertex $\la44444\ra$,
appears to give the same tilings.}

\subsection{Supertiling description}

\MEMO{See CLH queries in the Appendix.}
{\sl We can try to rationalize the emergent constraints by
considering the sum rules for the numbers of vertices
and other environments in a Rectangle-Triangle tiling,
as worked out in Appendix~\ref{app:sum-rules}.}

\SAVE{It is interesting that the HREM image tiling 
(Fig.\ref{fig:known-tilings}(c) 
is completely covered by SH$_2$ clusters
(with an outline of a decagon with two ears added;
sometimes we call this a ``lightbulb'' shape).
Note that many small approximants are built from SH$_2$ clusters, 
e.g. the 18-node tiling.
We conjecture the number of SH$_2$ clusters
is optimized by the 76-node tiling shown in
Figure ~\ref{fig:MC-tilings}(a).
This would be an example of a cluster maximization that
gives a large approximant crystal, in contrast to the
cluster maximization that can enforce a quasiperiodic 
ground state\cite{steinet98}.}

One property of the emergent ensemble
is that zigzag edges occur in chains with two
vertices of type $\la 443333 \ra$ and two endpoints
of type $\la 54443 \ra$, so that there are equal 
numbers of those two vertex types (see the discussion
of sum rules in Appendix~\ref{app:sum-rules}.)
A second salient property is that all Fat Rhombi 
occur in groups of five with a common tip, forming Star tiles.
\SAVE{A corollary is that the perimeters of the
Hexagons and Stars occupy only two levels of the
discrete part of the perp-space; the only other
vertices are the Star centers, which occupy the
outer two levels.  Thus the perp space symmetry
is like a Penrose tiling (centered between two levels)
and not a binary tiling (which has three occupied levels).}

\MEMO{The reconstructed tiling from the d-ZnMgDy HREM in 
Fig.\ref{fig:known-tilings}($c$) can be considered an HBS tiling.}

The only known quasiperiodic rectangle-triangle tiling
is Cockayne's iterative construction~\cite{cockayne}.
\SAVE{(The window function
(acceptance domain) describing an ideal quasiperiodic tiling
of this class is necessarily fractal~\cite{gaehler-fractal};
cite Kalugin?)}

That has superficial similarities to our energy optimal
tilings:  Cockayne's tiling is also entirely
built from Fat Rhombi and Hexagons, and it uses the
same five vertex types (boxed in Fig.~\ref{fig:vertices}).
But it contains few Stars -- instead, it has 
frequent clusters of five {\it Hexagons} sharing
a tip.  Also, the zigzag edges occur in chains with 
{\it one} vertex of type $\la 443333 \ra$, so
there are twice as many $\la 54443 \ra$ endpoint vertices 
as $\la 443333 \ra$ vertices. \SAVE{Finally, Cockayne's
tiling occupies three levels in perpendicular space with 
a symmetry around the middle one, like a binary tiling.}

Since the Star seems to be an important motif, we considered
marking the network of those points [Fig.~\ref{fig:MC-tilings}(b)].
We find they are typically separated by four kinds of ``linkage''
denoted $\alpha$, $\beta$, $\gamma$, and $\delta$.  
In contrast to more familiar such networks, only two of them 
lie in a symmetry direction ($\beta$ is parallel to the 
tile edges of length $a$, and $\alpha$ is parallel to 
the $b$ edges.)  The $\alpha$ linkage are drawn curved because
one side is inequivalent to the other. The $\beta$ linkage
have a mirror symmetry along the linkage while the $\alpha$
linkage has a mirror symmetry in the bisecting plane.  
Both $\gamma$ and $\delta$ edges have twofold symmetry 
around their centers. A $\beta$ edge can only be bordered
by a B tile, 
and similarly a $\gamma$ edge can only 
be bordered by a C tile, so these kinds come in pairs.
The $\alpha$ linkage connects vertices of the same parity
whereas the three other kinds connect vertices of opposite
parity.  By simple counting of the edges, a sum rule 
can be derived for the tile counts $N(B)=N(A)+N(C)$;
by considering their content as reduced (ultimately) to
Fat and Skinny Penrose rhombi, the numbers of each tile
are fully constrained (given a particular system cell,
or alternatively given the condition of full decagonal symmetry).

In this tiling, all vertices are of the five allowed types
(CHECK), and zigzags appear purely in chains
of two $\la 443333 \ra$ and two $\la 54443 \ra$ type vertices,
each of which runs down a pair of C tiles.

\MEMO{We have to check whether any such tilings can be 
built to fill the simulation cell!  If so, do they have
a somewhat better energy?  Also, I need to check the vertex
counts in this model.  I notices that tiling 29,
which was the imperfect one in terms of Marek's old 
analysis, is the perfect one in terms of CLH's new supertiling.}

It appears there is a nonzero entropy of ways to fill space
with these supertiles. However, their asymmetrical nature means
they are highly constrained so the entropy per node 
will be quite small, and correspondingly the vertices
will have position correlations over many multiples of 
the basic tile edge $a$.

\subsection{Crystallographic consequences}

\MEMO{This subsection and the next, logically, might belong
to a separate section about the results as an atomic model
(as opposed to a abstract tilings and supertilings).
Then the paper's structure would be that we went into
tiling-land, captured our treasure, and came out again
into the real world.  However, in the present draft,
the section seems too short to stand on its own.
What do you think?}

\MEMO{Tell the space group of the resulting decagonal structure.}

The fact that the Rectangle/Triangle tiling groups into
Hexagons and Stars has an important consequence for the 
atomic structure: following the rule for heights $z=0$ or 0.5 stated in 
Sec.~\ref{sec:decoration-atoms}, all mid-edge Zn atoms
on the borders of Hexagons or Stars have the same height,
globally; the only mid-edge Zn at the other height are
those decorating the five edges around each Star center.
(That also explains why Y atoms are observed only in
at one height.)
\SAVE{(MM in his 2001 draft pointed out this fact in the Roth-Henley model
``locks the ``up'' and ``down'' alternation such that the two levels 
are not statistically equivalent.''}

In the minimum-energy tilings, the Y sites (Sec.~\ref{sec-whereY}) 
sit in tips of Fat Rhombi that stick in between two Hexagons.
This pattern is always part of a decagon, consisting of 
a Hexagon plus three Fat Rhombi, that would be a regular
decagon except that one vertex is ``dented'' inwards.
The Y sites constitute the centers of all these
dented decagons, and (for the same reason as the mid-edge Zn sites)
all have the same $z$ value.  These two facts have been observed for
the Y sites in preliminary refinements of single-crystal data~\cite{steurer-solve}.

\subsection{Energetic stability}

Experimentally,  the phase stability around the composition
Mg$_{40}$Zn$_{60}$ at low $T$ is poorly studied, due to 
diverging equilibration times and further impeded by
the low melting point of Mg, and because 
the low formation enthalpy of the MgZn$_2$ Laves phase overshadows 
all other phases at low$T$.
\SAVE{(strong energetic dominance)}
\MEMO{``is it because
these phases have to form by solid state reaction of MgZn$_2$ with something?''}

\MEMO{WHICH tiling won? Is the following sentence true?}

{\sl None of these tilings turned out to be stable
but all tilings lacking R--R contacts are 
above the tie-line of competing phases
by just a marginal amount (a few meV).}

\MEMO{Is this the final conclusion?}
{\sl The Mg$_4$Zn$_7$ phase~\cite{mg4zn7} 
is the stable one~\cite{okamgzn} in our composition range
(practically on the tie-line at $T=0$, according to ab-initio calculations).
}

Say more about the approximant of~\cite{AT00}.

The addition of Y is found to make the quasicrystal or approximant
structures {\it stable} with respect to competing phases.
This may be attributed to two things
(1) the strong nearest neighbor Zn--Y attraction 
(see Fig.~\ref{fig:potentials}).
(2) the poor fit of Y in substituting into competing
structures.
\MEMO{Amplify on what Y does in the competing structures.}

\MEMO{MM: wants to say more about the
triangle-rich structures (which are the 
best candidates to be actually stable in
the Mg--Zn phase diagram).}

Notice that the \mgivznvii\ tiling is triangle-rich;

\section{Discussion}

In conclusion, 
we have reported a realistic study of structural stability and energetics 
at low temperature of the binary decagonal quasicrystal $d$-MgZn, and 
crystal structures that approximate it, and of the ternary $d$-MgZnY.
The binary $d$-MgZn was found 
\SAVE{(through ab-initio calculations)}
to be slightly unstable relative to competing Mg--Zn phases;
the inclusion of a small number of Y atoms tends to stabilize the quasicrystal,
as they substitute easily in certain kinds of Mg sites, but are less
easily accomodated in the competing, simpler crystal structures.

We computed the relaxed energy for every tiling in an 
enumeration of medium-sized tilings (using pair potentials) 
and then, by numerical fitting, 
re-expressed the total energies in terms
of a ``tiling Hamiltonian'' with single-tile energies as well
as tile-tile interactions~\cite{mihet96a}.

\SAVE{Our tile-decoration structural model of the d-MgZn 
quasicrystal, is more exact than that of Ref.~\cite{RH97}.
Our framework deviates from that of Ref.~\cite{RH97}
in being based on the ``rectangle-triangle'' tiling.}

Due to the large cost of adjoining rectangles,
our tiling is further constrained to consist of Fat Rhombi 
and Hexagons (which are respectively combinations of two T tiles, 
or of two T's plus one R); this simplifies the optimization
so much that there are essentially no free parameters in
the effective Hamiltonian, and the resulting tiling satisfies
highly constraining local rules.

\MEMO{to add: Summarize role of the RE}

\SAVE{In future work, we should also verify
the correctness of the basic decoration rule
using ``melt-quench'' MD simulations~\cite{MQ-AlCoNi,EOPP}.
That means the system is simply cooled from high temperatures, 
in a cell of properly chosen dimensions, 
so that an  ordered (quasicrystal) structure is 
obtained rather than a glass.}

\MEMO{(To SAVE?)[From MM 4/27/11]}
{\sl ``Vibrational entropy might stabilize a small atom on a large-atom site at $T>0$;
the candidate sites for this are those of coordination number CN=14, which are on the 
borderline between small-atom and large-atom size.''}
\MEMO{Was that referring to a recently discovered
decagonal phase stabilized by small atoms instead of RE?
If so CITE!}

\MEMO{(to SAVE?). CLH adds: Al--TM decagonals had ``channels'';
icosahedral Al--TM and Ca--Cd crystals have inner clusters
of reduced symmetry.  In either case, these are places in the
structure where stuctural uncertainty and fluctuations are concentrated.
By stabilizing some kind of superstructure, they may help 
establish long-range order in the the quasicrystal.
Is it possible that the specific placement of Y atoms can play such 
a role in FK decagonals?}

\MEMO{(to SAVE?) From MM 4/27/11:}
{\sl This study suggests the possibility of predicting quasicrystal phases in
many other alloy systems that have a Laves phase, when 
the content of large atoms is increased.}

\MEMO{Do we include this?}
Since Dy atoms are magnetic, our structure determination
also opens a door to accurate modeling of the antiferromagnetism
~\cite{Sa02} in these materials.  The RE sites we propose form
a network with nearest-neighbor spacing $\tau b\approx 8.5$\AA.~
\MEMO{CHECK}. It contains many pentagons, so it is non-bipartite
and (assuming the RKKY interaction at that distance is
antiferromagnetic) is frustrated.

Highly ordered {\it icosahedral}~\cite{TNIMNTT94} 
quasicrystals are long known in the ZnMgRE system, too.
(with composition Zn$_{64}$Mg$_{27}$Y$_9$\cite{AT99}, for example).
Now that the RE site has been identified, it would be
of interest to explain how RE stabilizes those structures,
and to examine the exact real-space structure.

\subsection[*]{Acknowledgements}
Marek - his slovak grants. PLEASE FILL IN.
MM, CLH, and J. R-D. by U.S. Dept. of Energy grant DE-FG02-89ER-45405.

\appendix

\section{Alternative redecorations}
\label{app:alternative-deco}

Here, we consider whether it was proper to limit consideration to
a rectangle/triangle tiling. Might another version of the tiling,
with different or additional tiles, permit structures of competitive
or even lower energies?  Before addressing this numerically, it
is helpful to consider the general problems of decoration models
with 10-fold symmetry.  The natural tiles are the Fat and Skinny
Penrose rhombi, respectively having acute angles $2\pi/5$ and $2\pi/10$.

The problem is that the Skinny rhombus is too thin.  
In the Skinny rhombus, the edge-decorating
atoms interfere sterically with the atoms on the 
opposite side of the short axis. There is no
freedom to adjust them: decorations along {\it all} 
tile edges are constrained to be the same,
so that any 
edge can fit with any other edge of the same type.
The only way to fix this problem is to modify the 
tiling so as to eliminate the (single) Skinny rhombus 
as a tile.
 
There are three alternative ways to do this.
First and most important, wherever the Skinny 
rhombi appear in pairs with two sharp tips adjoining, 
there is a Fat rhombus between the tips,
and the three can be combined as a Hexagon (H) tile. 
The edge shared by the two Skinny rhombi is now
part of the Hexagon's interior, and we are
allowed to redefine its interior decoration
while respecting its end-to-end mirror symmetry.
In most cases, the tips are decorated identically
to Fat Rhombi; then the most economical
description is to divide the Hexagon 
into rectangles and triangles for decoration
purposes (see Figure~\ref{fig:deco}).
Thus, {\it the Rectangle tile is a substitute
for thin rhombi.}

However, the Rectangle/Triangle tiling is not compatible 
with known kinds of matching rule or with any easy ways
of annealing the structure (either in simulations or
in reality).  Thus, we should check whether there are
other energetically acceptable ways to evade the presence
of Skinny rhombi: even if those are rare, they may have
a significant effect on the quasicrystal's properties.
\SAVE{
Another motivation is 
that a simple decoration of supertile rhombi with the 
basis tiles could be devised that would implement Penrose's 
arrow matching rules, if only a Fat Rhombus or Boat tile
were allowed.
(These are $\tau^2$ rhombi, so each Fat super-Rhombus
gets 5 Fat and 3 Skinny Rhombi, i.e. 1 R + 10 T + 1 leftover
skinny rhombus.)}

\begin{figure}
\caption{(a). Rhombus decoration of an isolated Skinny rhombus,
which forces Zn--Zn steric clashes; the Skinny rhombus may be 
combined with the rhombi above, forming a Boat tile, or with
the rhombi below, forming a Lozenge.
(b). Best redecorations of Boat
(c). Best redecorations of Lozenge tile.
}
\label{fig:alt-skinny}
\begin{center}
   \includegraphics[width=5.5in]{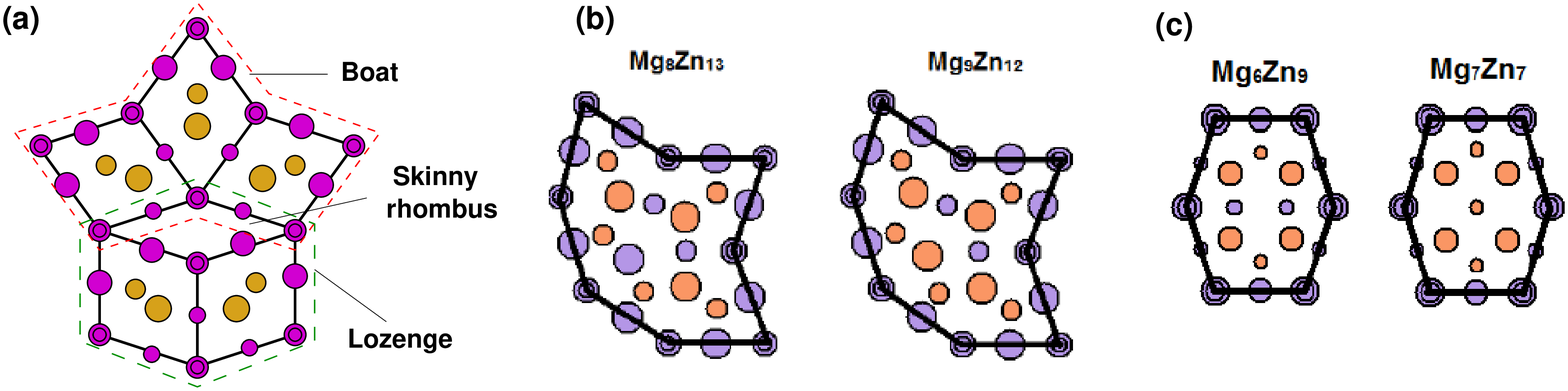}
\end{center}
\end{figure}

The Hexagon or Rectangle absorbs a {\it pair} of single Skinny rhombi.
The alternative is to combine a {\it single}  Skinny rhombus 
with adjacent Fat Rhombi in a composite tile.
Figure~\ref{fig:alt-skinny}(a) shows a typical environment of an
unpaired Skinny rhombus in a Penrose tiling; the four zinc atoms
on the two vertices at center are sterically interfering.
We may either combine the Skinny Rhombus with the three Fat Rhombi 
above it, forming the well-known Boat tile, or with the two Fat Rhombi
below, forming a sort of hexagon we will call a Lozenge tile.

In either case, the redecorations affect only the Zn atoms on
the Skinny rhombus vertex in the interior of the composite
tile, and the adjacent Zn atoms on edges.  We tried
(\MEMO{how many?}) candidate decorations, testing
their energies using a tiling in which they were
surrounded by Triangle tiles.  Fig. ~\ref{fig:alt-skinny}(b) 
shows the best two solutions we found for the Boat and
Fig. ~\ref{fig:alt-skinny}(c) shows the two best solutions for
the Lozenge.  In all cases, the energies were higher than
tilings of rectangles and triangles with the same atom content.

\MEMO{Need to be more complete in telling the energy costs.}


\MEMO{Alternative text}
{\sl In a rhombus tiling, the $144^\circ$ corner of an Skinny rhombus,
not adjoining another of its kind, must adjoin Fat rhombi:
either two $108^\circ$ corners, or else three $72^\circ$ acute corners;
in the Penrose tiling, indeed, the usual environment of an unpaired
Skinny rhombus has one such group on one side, and the other kind on
the other side.  Combining with one of these neighbor groups produces, 
respectively, a Fat Hexagon tile (which has more symmetry than
its constituents) or a Boat tile.  However, in the Mg--Zn binary,
all of these alternatives were found to be higher in energy.}

\SAVE{We could also go on to combine 
a Boat plus two Hexagons to form a Decagon 
(which can have much more symmetry than its constituents).}

\section{Sum rules on vertex counts}
\label{app:sum-rules}

An {\em arbitrary} R--T tiling (allowing all ten vertex types)
obeys two hidden constraints:
     \begin{subequations}
     \label{eqs:constraints}
     \begin{eqnarray}
     2 N(44444)+N(54443)=N(533333)
            \label{eq:C1}\\
     N(44444)=N(54533)
            \label{eq:C2}.
     \end{eqnarray}
     \end{subequations}
Eq.~\ref{eq:C2} can be understood by considering the network
of all $b$ bonds. First note that the number of $b$ bonds is equal
to the number of vertices (apportioned according to corner angles)
in both the Rectangle and the Triangle, and therefore in the
whole tiling.  Thus, the average coordination number of $b$ bonds,
$\bar{Z}_b$, must be exactly two.  The actual coordination number
$Z_b$ 3 for vertex $\la 533333$, one for vertex $\la 54443\ra$,
and zero for vertex $\la 44444 \ra$; the condition for $\bar{Z}_b=2$
is Eq.~\eqref{eq:C1}.

\MEMO{CLH doesn't see how you derive \eqref{eq:C2} from this.
Also, should derivations of sum rules be moved to an Appendix?}

Another kind of sum rule is derived by counting the number
of corners of each type at the vertices.  If we {\it assume}
only the five vertex types boxed in Figure~\ref{fig:vertices}
are present, this gives four independent equations on them:
\begin{subequations}
\label{eqs:rrt}
\begin{eqnarray}
N(44444) = N(54533) = 2N_R - N(54443) \label{eq:r1}\\
2N(443333) = N_T-6N_R             \label{eq:r2}\\
N(533333) = 2N_R\label{eq:r3}
\end{eqnarray}
\end{subequations}
Only one degree of freedom is left, which is the conversion
of a pair $\la 44444\ra + \la 54533 \ra \leftrightarrow 2\la 54443\ra$.

\MEMO{I'm confused to understand in what sense these
equations are independent.  I notice that \eqref{eq:r1} can
be derived from \eqref{eq:r3}, using \eqref{eq:C1}.}

Since $\la 54443 \ra$ has disfavorable ``zigzag'' bonds,
it is advantageous to minimize it.
\MEMO{CLH to MM: I didn't understand the logic in your 2001 paper,
for why the minimum of $N(54443)$ is equal to $N(443333)$,
these were $N_{41}$ and $N_{42}$ in your old notation.
You wrote ``each $V_{42}$ vertex forces at least one $V_{41}$ 
vertex nearby, the {\em minimum} of $N_{41}$
coincides with $N_{42}$.''
See the following argument, which you told me recently and
which seems to contradict the old assertion.}

A third kind of constraint
is obtained by imagining the 
network formed by ``zigzag'' type $a$ bonds, assuming only
the five boxed vertex types are present.  Of these, vertex
$\la 443333\ra$ contains {\it aligned} two zigzag bonds,
vertex $\la 54443\ra$
contains one, and the other vertices contain none.
Thus zigzag edges must form straight chains having $m\ge 0$ vertices
of type $\la 443333$, to endpoints of
type $\la 54443\ra$, and $(m+1)$ zigzags overall.

\MEMO{I'm confused at the arguments you made for why $m=1$
is optimal.  I thought you said that we were constrained
to a certain total number of zigzag edges, but the number
of endpoints was variable. And the endpoints $\la 54443$
were energetically costly.  Thus, you said, we had a trade-off:
long chains reduced the endpoint energy cost, but 
increased the perp-space distortion, so it is harder
to place the fivefold symmetric $\la 44444\ra$ vertex.
(Which in turn means there must be more $\la 54443$ vertices,
since those are traded for $\la 44444\ra$.)  However, it seems
to me the only reason that $\la 54443\ra$ is costly is 
because it contains a zigzag, and if that number is fixed
then it shouldn't matter`whether they appear in $\la 54443\ra$
vertices or in $\la 443333\ra$ vertices.}

\section{The entire Mg--Zn phase diagram}
\label{sec:grand-Mg-Zn}

\MEMO{This is the lead-in typically used by MM.
CLH felt it was too much detail for the introduction.
Just parked here while we decide what to do with it.}

The binary Mg--Zn alloy includes stable (or nearly stable) 
compounds related
to {\it four} kinds of quasicrystal order:
\begin{itemize}
\item[(1)]
Mg$_2$Zn$_{11}$ contains a variant of the Tsai cluster 
seen in the Ca--Cd class of quasicrystals 
(lacking the outermost shell);
\MEMO{Is that the ``pseudo-Tsai'' cluster also found in
$oC104$-Al$_{38.8}$Cu$_{45.7}$Sc$_{15.5}$, \cite{EOPP}?}
\item[(2)]
layered structures related to Frank-Kasper (FK) decagonals,
\SAVE{(Mg$_{21}$Zn$_{25}$ is also a defective FK [decagonal??] structure)}
\item[(3)]
orthorhombic Mg$_{51}$Zn$_{20}$ (sometimes called
Mg$_7$Zn$_3$) 
a bcc packing of the ``Mackay icosahedron'' clusters 
familiar from Al--transition metal quasicrystals
\SAVE{(i.e. a ``1/1'' approximant)}
\SAVE{The first three categories are stable; ico FK 
and the rest are $barely$ unstable.}
\item[(4)]
icosahedral FK structures around 40\%Mg content.
\SAVE{(In the composition range 0.33$<x_{Mg}<$0.43, all
the low-energy Mg--Zn structures are related to FK
quasicrystals, of either the icosahedral or decagonal kind.}
\end{itemize}
Besides this, one also finds $\beta$-Al$_3$Mg$_2$ order also around 40\% Mg
a binary variant of the recently refined $\phi$--AlMgZn with the exact formula
MgZn, and the Al$_{30}$Mg$_{23}$ structure.
Some slightly unstable phases might become stabilized by
addition of a third species, e.g. decagonal MgZnRE with only 
2\% of RE (rare earth) atoms, or 2--9\% of Al for several newly 
discovered phases in the Al--Mg--Zn ternary system~\cite{berthold-kreiner}.


\begin{thebibliography}{11}

\bibitem{FK} Frank-Kasper

\bibitem{hafner}
J. Hafner,
{\it From Hamiltonians to phase diagrams :
the electronic and statistical-mechanical theory of
sp-bonded metals and alloys}
(Springer,  Berlin, 1987)

\bibitem{RH97}
J.~Roth and C.~L. Henley,
  {Phil. Mag. A} \textbf{75}, {861} (1997).

\bibitem{Ni94}
A. Niikura, A.~P.~Tsai, A.Inoue, and T.~Masumoto,
Phil Mag Lett 69, 351 (1994).
TSAI AP; INOUE A; et al.

\bibitem{Tsai94b}
A. P. Tsai, A. Niikura, A. Inoue, T.Masumoto,
Y. Nishida, K.~Tsuda, and M.~Tanaka,
Phil. Mag. Lett 70, 169 (1994)


\bibitem{A98}
E. Abe,  T.~J.~Sato, and A.~P.~Tsai
Philos. Mag. Lett. 77, 205 (1998)

\bibitem{AT99}
E.~Abe and A.~P. Tsai,
  {Phys. Rev. Lett.} \textbf{83}, {753} (1999).

\bibitem{suck-meltspun}
R. Sterzel, E. Dahlmann, W. Assmus, K. Saitoh, H. Fuess, M. Mihalkovic 
and J.-B. Suck, Phil. Mag. Letters, 82 (4), 235
TO CITE THIS

\bibitem{steurer-solve}.
T. \"{O}rs and W. Steurer,
personal communication.

\bibitem{steurer-grow}
T. \"{O}rs and W. Steurer, 
Phil. Mag. 91, 2466 (2011)

\bibitem{vasp}
(a) G. Kresse and J. Hafner, 
Phys.\ Rev. B, 47, R558 (1993);
(b) G. Kresse and J. Furthmuller,
Phys. Rev. B 54, 11169 (1996).

\bibitem{TNIMNTT94}
A.~P. Tsai, A.~Niikura, A.~Inoue, A.~Masumoto,
  Y.~Nishida, K.~Tsuda, and M.~Tanaka,
  {Phil. Mag. Lett.} \textbf{70}, {169} (1994).

\bibitem{SAT98}
T.~J. Sato, E.~Abe, and A.~P. Tsai,
  {Phil. Mag. Lett.} \textbf{77}, {213} (1998).

\bibitem{AT00}
E.~Abe and A.~P. Tsai,
  {Acta Cryst. B} \textbf{56}, {915} (2000).

\bibitem{AST99}
E.~Abe, T.~J. Sato, and A.~P. Tsai,
  {Phys. Rev. Lett.} \textbf{82}, {5269} (1999).

\bibitem{Abe-HREM} some of the Abe papers, but which?


\bibitem{EOPP}
M. Mihalkovi\v{c} and C.~L.~Henley,
preprint (arXiv:1109.6931),
``Empirical oscillating potentials for alloys from ab-initio fits,
and the prediction of quasicrystal-related structures in the Al-Cu-Sc system''

\bibitem{friedel-FK}
J. Friedel, Helv. Physica Acta 61, 538 (1988).


\bibitem{Lim08a}
S. Lim, M. Mihalkovi\v{c}, and C. L. Henley,
Philos. Mag. 88, 1977-1984 (2008).

\bibitem{Lim08b}
S. Lim, M. Mihalkovi\v{c}, and C. L. Henley,
Z. Kristallogr. 223, 843-846 (2008).

\bibitem{OxHen93}
M. Oxborrow and C.~L.~Henley, 
Phys. Rev. B 48, 6966-6998 (1993).

\bibitem{OM95}
M.~Oxborrow and M.~Mihalkovi\v{c}, in
  {\it Aperiodic '94, Proceedings of the International
  Conference on Aperiodic Crystals}, edited by
  G.~Chapuis and W.~Paciorek
  (World Scientific, 1995), p. {178}.

\bibitem{OM98}
M.~Oxborrow and M.~Mihalkovi\v{c}, in
 {\it Aperiodic 97, Proceedings of the International
  Conference on Aperiodic Crystals}, edited by
  M.~de~Boissieu, J.-L. Verger-Gaugry, and R.~Currat
  (World Scientific, 1998), p.  {451}.

\bibitem{cockayne}
E.~Cockayne, {Phys. Rev. B} \textbf{51}, {14958} (1995).

\bibitem{mihet96a}
M. Mihalkovi\v{c}, W.-J. Zhu, C. L. Henley, and M. Oxborrow,
Phys. Rev. B 53, 9002 (1996).
\MEMO{DO WE WANT TO CITE DECO II?}

\bibitem{mg4zn7} 
experiment on Mg$_4$Zn$_7$, 1990s?

\bibitem{okamgzn} 
``Okamoto, MgZn, circa 1998?'' \MEMO{Is this it??}
H. Okamoto,"Comment on Mg-Zn (Magnesium-
Zinc)", J. Phase Equilib. 15 (1994) 129-130.


\bibitem{MMITDMMMHY93}
U.~Mizutani, Y.~Yamada, T.~Takeuchi, K.~Hashimoto,
  E.~Belin, A.~Sadoc, T.~Yamauchi, and T.~Matsuda,
  {J. Non-Cryst. Sol.} \textbf{156--158}, {882} (1993).
UNCITED

\bibitem{MM95}
M.~Mihalkovi\v{c}, in
{\it Aperiodic '94, Proceedings of the International
  Conference on Aperiodic Crystals}, edited by
  G.~Chapuis and W.~Paciorek
  (World Scientific, 1995), p. {552}.
UNCITED

\bibitem{YKM75}
Y.~P.~Yarmolyuk, P.~I.~Krypiakevich, and E.~V.~Me\'lnik, 
Sov. Phys. Cryst.  \textbf{25}, 329 (1975).
NEED TO CITE?




\bibitem{mihet02}
M.~Mihalkovi\v{c}, I. Al-Lehyani, E.~Cockayne,
C.~L.~Henley, N.~Moghadam, J.~A.~Moriarty, Y.~Wang, and M.~Widom,
Phys. Rev. B 65, 104205 (2002).
UNCITED

\bibitem{Sa02}
T. J. Sato, H. Takakura, J. Guo, A. P. Tsai, and
K. Ohoyama,
J. All. Compounds 342, 365-8 (2002).

\bibitem{steinet98}
P.~J. Steinhardt, H.-C. Jeong, K.~Saitoh, M.~Tanaka, E.~Abe, and A.~P. Tsai,
  {Nature} \textbf{396}, {55} (1998).

\bibitem{berthold-kreiner} 
Berthold and G. Kreiner (TO FILL IN)


\SAVE{\bibitem{berg52}
G.~Bergman,
Acta Cryst.~\textbf{10}, 254 (1952).}

\SAVE{\bibitem{kreiner02}
G. Kreiner, J. Alloys and Comp. 338, 261 (2002).}

\SAVE{
\bibitem{Roth95} 
J.~Roth, R.~Schilling, and H.-R. Trebin,
Phys. Rev. B51, 15833 (1995).
}

\SAVE{
\bibitem{gaehler-fractal}
F. G\"ahler, doctoral thesis, No. 8414,
ETH Z\"urich (1988);
M. Baake, R.Klitzing, and M.~Schlottmann,
Physica A 191, 554 (1992);
A. P. Smith,
J. Non-Cryst.~Solids 153\& 154, 258 (1993).
}

\bibitem{MQ-AlCoNi}
M. Mihalkovi\v{c}, C.~L.~Henley, and M.~Widom,
Philos. Mag.  91,   2557-2566 (2011).

\OMIT{
\bibitem{widom-GPT} 
J. A. Moriarty and M. Widom,
Phys. Rev. B 56, 7905 (1997).}





\end{thebibliography}
\end{document}